\documentstyle[aps,prl,multicol,epsfig]{revtex}
\newcommand{\bq}{\begin{equation}}
\newcommand{\eq}{\end{equation}}
\newcommand{\bqa}{\begin{eqnarray}}
\newcommand{\eqa}{\end{eqnarray}}
\newcommand{\nn}{\nonumber \\}

\begin{document}
\draft
\title{Universal scaling behavior of pseudogap with doping in high $T_c$ cuprates; temperature and doping dependence of spectral intensity} 

\author{Sung-Sik Lee and Sung-Ho Suck Salk$^a$}
\address{Department of Physics, Pohang University of Science and Technology,\\
Pohang, Kyoungbuk, Korea 790-784\\
$^a$ Korea Institute of Advanced Studies, Seoul 130-012, Korea\\}
\date{\today}

\maketitle

\begin{abstract}
Based on our improved SU(2) slave-boson approach (Phys. Rev. B {\bf 64}, 052501(2001)) to the t-J Hamiltonian, we report a scaling behavior of pseudogap with doping and the temperature and doping dependence of spectral functions.
In addition we discuss the cause of hump and quasi-particle peak in the observed spectral functions of high $T_c$ cuprates. 
It is demonstrated that the sharpening of the observed quasi-particle peak below $T_c$ is attributed to the bose condensation of holon pair. 
From the computed ratios of pseudogap $\Delta_0$ to both the superconducting temperature $T_c$ and the pseudogap temperature $T^*$ as a function of hole doping concentration $x$, we find that there exists a universal scaling of these ratios with doping, that is, the hyperbolic scaling behavior of the former, $\frac{2 \Delta_0}{k_B T_c} \sim x^{-\alpha}$ with $\alpha \sim 2$ and near doping independence of the latter, $\frac{2 \Delta_0}{k_B T^*} \approx 4 \sim 6$ are found.
\end{abstract}
\begin{multicols}{2}

\newpage
Angle resolved photoemission spectroscopy(ARPES) measurements of high $T_c$ cuprates revealed both the temperature and doping dependence of both spectral peak intensity and pseudogap at varying pseudogap temperature $T^*$ and superconducting temperature $T_c$.
These measurements have shown a continuously increasing trend of pseudogap (leading edge gap) with decreasing temperature even below $T_c$ and appearance of sharp peaks below $T_c$\cite{SHEN95,NORMAN}. 
Earlier, Wen and Lee\cite{WEN} reported the momentum dependence of the spectral function based on their SU(2) slave boson theory involving single-holon bose condensation.
Chubukov et al. obtained the peak-dip-hump feature of the spectral function by using their spin fermion model\cite{CHUBUKOV}.
Most recently, Muthukumar et al. studied the momentum dependence of the spectral function in the resonating-valence-bond state\cite{MUTH}. 
However there exists lack of comprehensive investigations on the temperature and doping dependence of spectral functions.
Lately, based on our newly improved SU(2) slave-boson approach of introducing holon-pair channels\cite{SALK} over a previous one\cite{GIMM} we were able to successfully obtain the arch-shaped superconducting temperature line in the phase diagram of high $T_c$ cuprates.
Further application of this theory resulted in  the peak-dip-hump structure of optical conductivity in good agreement with observations\cite{LEE_OPT}.
In the present study using this theory we report temperature and doping dependence of spectral functions for both the normal and superconducting states and discuss the origin of the hump and the quasiparticle peak observed in ARPES\cite{NORMAN}\cite{SHEN}. 
Finally we report our theoretical finding of the scaling behavior of the spin gap vs the superconducting and pseudogap temperatures with respect to the hole doping concentration.

In the slave-boson representation\cite{WEN}\cite{KOTLIAR}, the electron operator of spin $\sigma$, $c_{\sigma}$ can be written as a composite of spinon and holon operators; $c_{\sigma}  =  b^\dagger f_{\sigma}$ in the U(1) representation and 
$c_{\alpha}  = \frac{1}{\sqrt{2}} h^\dagger \psi_{\alpha}$ in the SU(2) theory with $\alpha=1,2$,
where $f_{\sigma}$($b$) is the spinon(holon) annihilation operator in the U(1) theory,
and $\psi_1=\left( \begin{array}{c} f_1 \\ f_2^\dagger \end{array} \right)$ $\left(\psi_2 = \left( \begin{array}{c} f_2 \\ -f_1^\dagger \end{array} \right) \right)$ and $h = \left( \begin{array}{c} b_{1} \\ b_{2} \end{array} \right)$ are the doublets of spinon and holon annihilation operators respectively in the SU(2) theory. 
Judging from the presence of the intersite charge-charge coupling ($n_i n_j$) in addition to the spin-spin coupling (${\bf S}_{i} \cdot {\bf S}_{j}$) in Heisenberg term of the t-J Hamiltonian and the consideration of on-site charge fluctuations\cite{CF}, it is obvious that the charge degree of freedom has to appear in the slave-boson representation of the Heisenberg term. 
The Heisenberg term in the t-J Hamiltonian is, then, derived to be $P({\bf S}_{i} \cdot {\bf S}_{j} - \frac{1}{4}n_{i}n_{j})P = - \frac{1}{2} b_i b_j b_j^\dagger b_i^\dagger (f_{\downarrow i}^{\dagger}f_{\uparrow j}^{\dagger}-f_{\uparrow i}^ {\dagger}f_{\downarrow j}^{\dagger})(f_{\uparrow j}f_{\downarrow i}-f_{\downarrow j}f_{\uparrow i}) $ in the U(1) theory, and  $- \frac{1}{2} ( 1 - h_{i}^\dagger h_{i} ) ( 1 - h_{j}^\dagger h_{j} ) (f_{2i}^{\dagger}f_{1j}^{\dagger}-f_{1i}^ {\dagger}f_{2j}^{\dagger})(f_{1j}f_{2i}-f_{2j} f_{1i})$ in the SU(2) theory \cite{SALK}.
Here $P$ represents the projection onto the singly occupied site or empty site.
The four holon (boson) operator in the above expression represents the charge degree of freedom and the four fermion operator, the spin degree of freedom.
In the present SU(2) study the phase fluctuation effects of the spinon pairing order parameter are taken into account.

Introducing Hubbard Stratonovich transformations for direct, exchange and pairing channels and a subsequent saddle point approximation, the t-J Hamiltonian is decomposed into the spinon sector, $H^f$ and the holon sector, $H^b$\cite{SALK},
\bqa
H^f  & = & -\frac{J(1-x)^2}{2} \sum_{<i,j>} \Bigl[ \Delta_{ij}^{f*} (f_{1j}f_{2i}-f_{2j}f_{1i}) + c.c. \Bigr] \nn
&& - \frac{J(1-x)^2}{4} \sum_{<i,j>,\sigma} \Bigl[ \chi_{ij} (f_{\sigma i}^{\dagger}f_{\sigma j}) + c.c. \Bigr], \nn
H^b &  = & -\frac{t}{2} \sum_{<i,j>} \Bigl[ \chi_{ij}(b_{1i}^{\dagger}b_{1j} - b_{2j}^{\dagger}b_{2i}) \nn
&& -\Delta^f_{ij} (b_{1j}^{\dagger}b_{2i} + b_{1i}^{\dagger}b_{2j})\Bigr] - c.c. - \sum_{i,\alpha} \mu_i b_{\alpha i}^\dagger b_{\alpha i} \nn
&& - \sum_{<i,j>,\alpha,\beta} \frac{J}{2}|\Delta^f_{ij}|^2 \Bigl[ \Delta_{ij;\alpha \beta }^{b*} (b_{\alpha i}b_{\beta j}) + c.c. \Bigr],
\label{eq:hamiltonian}
\eqa
where $\chi_{ij}= < f_{\sigma j}^{\dagger}f_{\sigma i} + \frac{2t}{J(1-x)^2} (b_{1j}^{\dagger}b_{1i} - b_{2i}^\dagger b_{2j} )>$ is the hopping order parameter, $\Delta_{ij}^{f}=< f_{1j}f_{2i}-f_{2j}f_{1i} >$, the spinon pairing order parameter, $\Delta_{ij;\alpha\beta}^{b} = <b_{i\alpha}b_{\beta j}>$, the holon pairing order parameter, $\mu_i$, the effective chemical potential, and $x$, the hole doping concentration.
With the uniform hopping order parameter, $\chi_{ij}=\chi$, the d-wave spinon pairing order parameter, $ \Delta_{ij}^{f}=\pm \Delta_f$ with the sign $+(-)$ for the nearest neighbor link parallel to $\hat x$ ($\hat y$) and the s-wave holon pairing order parameter, $\Delta_{ij;\alpha \beta}^{b}=\Delta_b(\delta_{\alpha,1}\delta_{\beta,1} - \delta_{\alpha,2}\delta_{\beta,2} )$, the quasiparticle energy for spinon is given by\cite{SALK}
\bqa
E_{k}^{f}  & = &  \sqrt{(\epsilon_{k}^{f})^{2} + ( \Delta_f^{'})^2},
\label{eq:spinon_energy}
\eqa
where the spinon single particle energy is given by,
\bqa
\epsilon_{k}^{f} & = & -\frac{J(1-x)^2}{2} \chi (\cos k_x + \cos k_y),
\label{eq:spinon_energy_single}
\eqa
and the spinon pairing gap or simply the spin gap,
\bqa
\Delta_f^{'} & = & J (1-x)^2 \Delta_f (\cos k_x - \cos k_y).
\label{eq:spinon_energy_gap}
\eqa
The single particle(electron) propagator of interest is given by a convolution integral of spinon and holon propagators in the momentum space\cite{WEN},
\bqa
G_{ \alpha \beta}({\bf k},\omega) & = & \frac{i}{2} \int \frac{d{\bf k}^{'} d\omega^{'}}{(2\pi)^3} \Bigl[ \sum_{l,m}G^f_{ \alpha \beta l m }({\bf k}+{\bf k}^{'},\omega+\omega^{'}) \times \nn
&& G^b_{ml}({\bf k}^{'},\omega^{'}) \Bigr].
\label{eq:convolution}
\eqa
The mean field Green's functions are $G^f_{\alpha\beta l m}({\bf k},\omega) = -i\int dt \sum_x e^{i\omega t-i {\bf k} \cdot {\bf x}} < T [ \psi_{\alpha l}({\bf x},t) \psi_{\beta m}^\dagger(0,0) ] >$
and $G^b_{lm}({\bf k},\omega) = -i\int dt \sum_x e^{i\omega t-i {\bf k} \cdot {\bf x}} < T [ b_{l}({\bf x},t) b_{m}^\dagger(0,0) ] >$ respectively.
The symbol $<$ $>$ refers to the finite temperature  ensemble average of an observable quantity $O$, $<O> \equiv \frac{1}{Z}{\rm tr}(e^{-\beta H}O)$.

The one electron removal spectral function, $I({\bf k}, \omega )$ is obtained from \cite{RANDERIA},
\bqa
I({\bf k},\omega) = -\frac{1}{\pi}{\rm Im G}({\bf k},\omega + i 0^+) f( \omega ), 
\eqa
where $f(\omega)$ is the Fermi distribution function.
The Heisenberg coupling constant of $J=0.2$ $t$ and the hopping strength of $t=0.44$ $eV$\cite{HYBERTSEN} are chosen in the present calculations.
Using the SU(2) theory, the predicted values of optimal hole doping $x_o$, pseudogap temperature $T^*$ and bose condensation temperature $T_c$ are $x_o = 0.13$, $T^* = 0.029 t$($148 K$) and $T_c = 0.021 t$ ($107.2 K$) respectively.
The spectral function is obtained from the convolution integral of the holon and spinon Green's functions $G({\bf k}, \omega)$.
For the evaluation of the spectral functions $I({\bf k}, \omega)$, the line width of the Lorentzian function is set to be $\epsilon = 0.01 t (4.4 meV)$ with the choice of $t=0.44 eV$.

It is known from the ARPES measurements of LSCO\cite{INO} and BSCCO\cite{DING97} that a good Fermi nesting appears in a region near the $M$ point (${\bf k}=(\pi,0)$) at which maximal antiferromagnetic correlations between electrons are realized.
The predicted binding energy monotonically decreases with the increase of momentum ${\bf k}$ along $\Gamma - M$, in agreement with the ARPES\cite{CAMPUZANO99}.
In Fig.1, predicted spectral functions at ${\bf k}=(\pi/2,\pi/2)$ below and above $T_c$ are displayed.
The peak did not disappear even at a temperature above $T_c$ (e.g., $T=0.022t$($112.3K$) with a choice of $t=0.44eV$) and showed a decreasing trend of peak intensity with increasing temperature,  in agreement with the ARPES measurement of Shen et al\cite{SHEN95}.
At all temperatures below $T^*$ the zero gap is predicted.

In Fig.2 we display the computed spectral functions $I({\bf k},\omega)$ at ${\bf k}=(\pi,0)$ both below and above $T_c$.
The computed gap size at ${\bf k}=(\pi,0)$ is predicted to coincide with the spin gap size $\Delta_f^{'}$ in Eq.(\ref{eq:spinon_energy_gap}).
The peak position occurs at smaller binding energies at temperatures ($0.022t$($112.3 K$)) above $T_c$. 
Although not shown here, a gap begins to open at higher critical temperature $T^* = 0.029t$($147.9 K$). 
This is now taken to be the spin gap temperature.
The predicted peak position shift, that is, the variation of leading edge gap is much slower below $T_c$ and reaches a finite value at $T = 0$, in complete agreement with observations\cite{CAMPUZANO99}\cite{DING}.

In order to see only the role of the antiferromagnetic fluctuations of the shortest possible correlation length, namely the spin singlet pair excitations in the evaluation of the convolution integral, we deliberately removed the contribution of the holon-pair channels which comes from momenta ${\bf k}^{'}=(0,0)$ and ${\bf k}^{'}=( \pi,\pi)$.
The sharp peak is now seen to disappear and only the hump structure persists while the pseudogap remains unchanged, indicating that the pseudogap is caused by the formation of spin singlet pairs as is shown in the inset of Fig.2.
This indicates that the presence of holon-pair bosons is essential for yielding the observed sharp quasiparticle peaks at the $M$ point below $T_c$. That is, the quasiparticle peak is caused by bose condensation below $T_c$. 
The appearance of the hump in the absence of bose condensation is seen to be caused by the antiferromagnetic spin fluctuations of  the shortest possible correlation length, that is, the spin singlet pair excitations.
In Fig. 3 the doping dependence of spectral functions is displayed for underdoped, optimally doped and overdoped cases at a low temperature below $T_c$, $T = 0.004 t$ ($T = 20.4 K$), near the $M$ point (${\bf k} = (0. 8 \pi, 0)$) for comparison with observation.  
In agreement with the ARPES\cite{SHEN}\cite{HARRIS} the predicted spectral weight of the sharp quasiparticle peaks is seen to increase as the hole concentration increases upto a tested value in the overdoped region. 
The leading edge gap is shown to decrease with doping in agreement with observations\cite{INO}.
At all doping rates both the peak and hump structures are revealed with the absence of dip.
It is observed to be caused by electron-phonon coupling\cite{LANZARA}, which is not considered in our present treatment of the t-J Hamiltonian.  
Figs.4 displays as a function of doping rate the ratios of the pseudogap $2 \Delta_0(k=(\pi,0))$ at a temperature close to $0 K$ to the superconducting temperature $T_c$ and the spin gap temperature $T^*$, that is, $2\Delta_0$ vs $T_c$ and $2\Delta_0$ vs $T^*$.  It is possible that each high $T_c$ cuprate(e.g., LSCO, YBCO and BSCCO) may have an effectively different $J$ value causing variation in $T_c$.
It is quite encouraging to find that in agreement with observations\cite{ODA}, $\frac{2 \Delta_0(k=(\pi,0))}{k_B T_c}$ decreases rapidly, showing a hyperbolic behavior with doping concentration while $\frac{2 \Delta_0(k=(\pi,0))}{k_B T^*}$ shows a nearly doping independence of ranging between $4$ and $6$, as shown in Fig. 4.
It is found that these ratios do not appreciably change with the variation of Heisenberg coupling $J$.
This is due to the nature that $\Delta_0$, $T^*$ and $T_c$ are found to be proportional to $J$, keeping these ratios unchanged\cite{SALK}.
Thus we found that independent of Heisenberg exchange coupling there exists a universal scaling behavior of the pseudogap with doping for the ratios of $\Delta_0$ to $T_c$ and $\Delta_0$ to $T^*$.
 
Both the SU(2) and U(1) slave-boson theories predicts sharp quasiparticle peaks at ${\bf k} = (\pi, 0)$ and ${\bf k} = (\pi/2, \pi/2)$ particularly below $T_c$.
Although not shown here, the predicted leading edge gap at ${\bf k}=(\pi,0)$ showed a continuous increase as temperature drops down from a pseudogap temperature $T^*$ to temperatures below the superconducting temperature $T_c$. 
This indicates that the origin of the observed leading edge gap in the superconducting state is the same as the ones in the pseudogap phase.  
That is, the pseudogap is caused by the formation of spin singlet pairs(spinon pairs).
The presence of the singlet pairs, namely the antiferromagnetic fluctuations of a shortest possible range is shown to persist even in the superconducting state.
We infer from our computed results that the appearance of the distinctively sharp quasiparticle peaks below $T_c$ is attributed to the enhanced probability of spin singlet pair excitations as a result of coupling to holon-pair bosons in the superconducting state.
It is of note that such coupling between the two degrees of freedom is well manifested in the last term of Eq.(\ref{eq:hamiltonian}).

In summary the present study revealed numerous salient features. 
They are;
1. regarding the hole doping dependence of spectral functions near ${\bf k}=(\pi,0)$, the spectral intensity of quasiparticle peak shows an increasing trend with hole concentrations in the underdoped region,
2. regarding the temperature dependence of pseudogap(spin gap) at ${\bf k}=(\pi,0)$, the gap size continuously increases as temperature decreases from the pseudogap temperature $T^*$ to temperatures even below the superconducting temperature $T_c$,
3. regarding the temperature dependence of peak intensity, a decreasing trend of spectral peak intensity is observed as temperature increases,
4. the zero leading edge gap at ${\bf k}=(\pi/2,\pi/2)$ is unchanged both below and above $T_c$, 
5. regarding the momentum dependence of binding energy, it decreases with increasing momentum ${\bf k}$ along the $\Gamma - M$ direction,
6. the sharp quasiparticle peak below $T_c$ is attributed to the bose condensation of holon pairs, and the hump is caused by the antiferromagnetic spin fluctuations of the shortest possible correlation length, 
7. the coexistence of the spin singlet pair excitations and superconductivity below $T_c$ is predicted, and finally,
8. in complete agreement with observations, independent of the Heisenberg exchange coupling there exists a universal scaling behavior of the pseudogap with doping, namely the hyperbolic scaling behavior in $\frac{2 \Delta_0}{k_B T_c}$ and near doping independence in $\frac{2 \Delta_0}{k_B T^*}$.

One(SHSS) of us acknowledges the generous supports of Korea Ministry of Education(HakJin Program 2001-2002) and the Institute of Basic Science Research (2002) at Pohang University of Science and Technology. 
He is also grateful to Chang Ryong Kim for helpful discussions.

%%%%%%%%%%%%%%%%%TWO COLUMNS%%%%%%%%%%%%%%%%%%%%%%%%%%%%%%%%%%%

\begin{minipage}[c]{9cm}
\begin{figure}
\vspace{0cm}
\epsfig{file=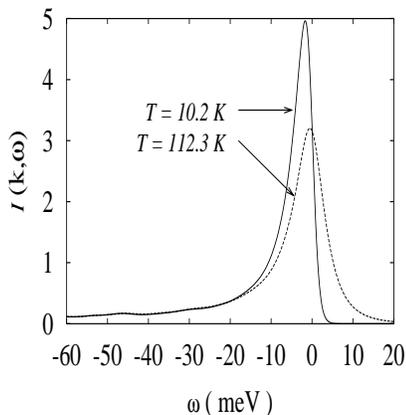, height=5.8cm, width=5.8cm}
\label{fig:1}
\caption{
Spectral functions $I({\bf  k}, \omega)$ for ${\bf k}=(\pi/2,\pi/2)$ at the predicted optimal doping of $x_o = 0.13$ ($T_c=0.021t$($107.2 K$), $T^*= 0.029t$($147.9 K$)) with $J=0.2t$ at $T= 0.002t$($10.2K$) below $T_c$ and $T=0.022t$($112.3K$) above $T_c$.
}
\end{figure}
 \end{minipage}

\begin{minipage}[c]{9cm}
\begin{figure}
\vspace{0cm}
\epsfig{file=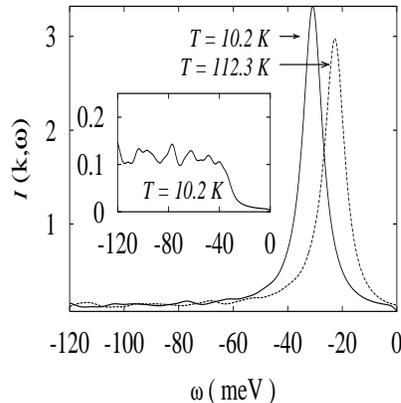, height=5.8cm, width=5.8cm}
\label{fig:2}
\caption{
Spectral functions $I({\bf  k}, \omega)$ for ${\bf k}=(\pi,0)$ at the same predicted optimal doping $x_o$, critical temperature $T_c$ and pseudogap temperature $T^*$ at the same temperatures below $T_c$ and above $T_c$ as shown in Fig.1.
The inset is $I({\bf k},\omega)$ for ${\bf k}=(\pi,0)$  with the exclusion  of  only ${\bf k}^{'}=(0,0)$  and  $(\pi,\pi)$  at $T=0.002t$($10.2 K$) below $T_c$.
}
\end{figure}
 \end{minipage}

\begin{minipage}[c]{9cm}
\begin{figure}
\vspace{0cm}
\epsfig{file=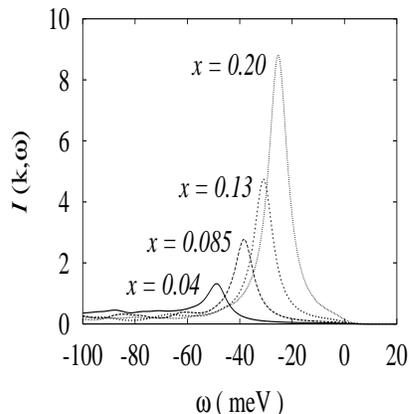, height=5.8cm, width=5.8cm}
\label{fig:3}
\caption{
Doping dependence of spectral functions $I({\bf k},\omega)$ near the $M$ point (${\bf k}=(0.8 \pi , 0)$) at temperature $0.004t$($20.4K$) below $T_c$ for underdoped($x=0.04$ and $x=0.085$), optimal doping($x_o =0.13$) and overdoped($x=0.2$) rates.
}
\end{figure}
 \end{minipage}

\begin{minipage}[c]{9cm}
\begin{figure}
\vspace{0cm}
\epsfig{file=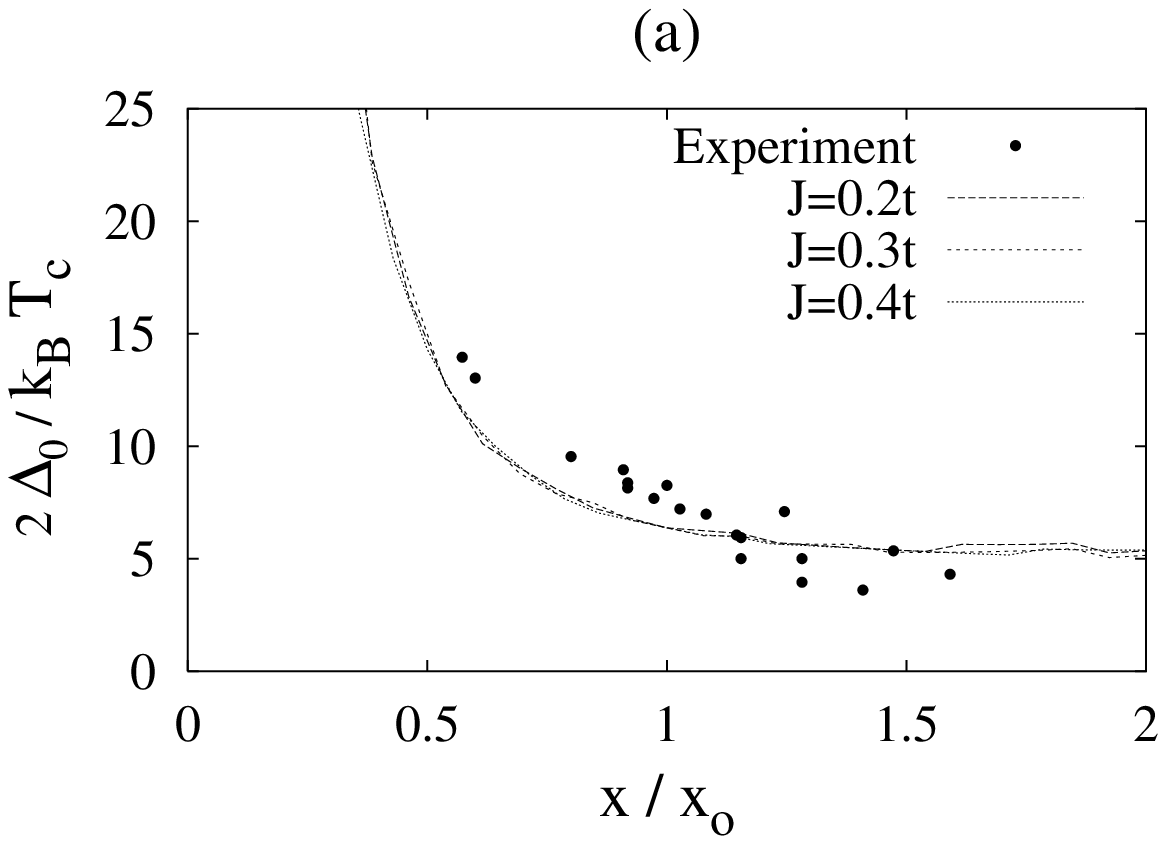, height=5.8cm, width=5.8cm}
\epsfig{file=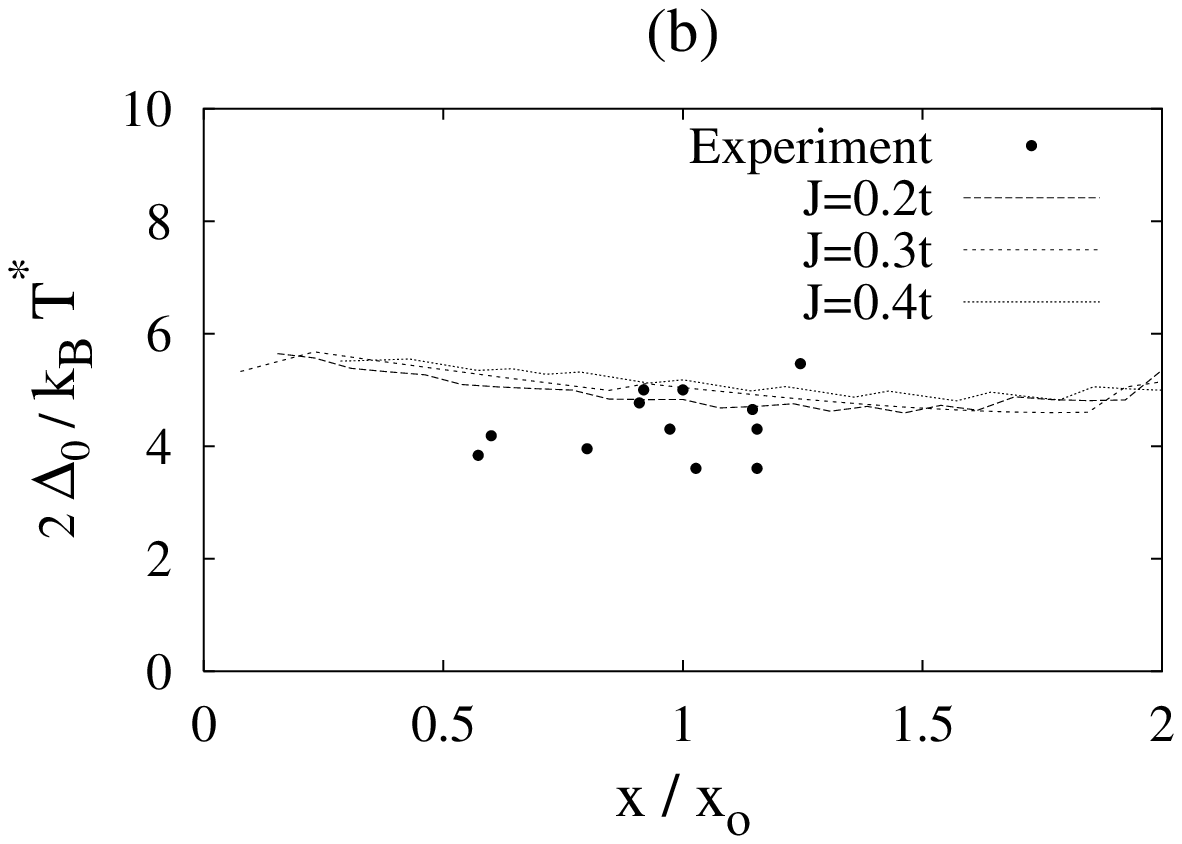, height=5.8cm, width=5.8cm}
\label{fig:4}
\caption{
The ratios of the spinon pairing gap at $T=0.001t$ ($\sim 5 K$) with the choice of $t=0.44 eV$[13] to 
(a) the superconducting temperature, $\frac{2 \Delta_0(k=(\pi,0))}{k_B T_c}$ and
(b) the spin gap temperature, $\frac{2 \Delta_0(k=(\pi,0))}{k_B T^*}$ 
as a function of $\frac{x}{x_o}$ where $x_o = 0.13$ is the predicted optimal doping concentration for $J/t=0.2$ and $0.3$, and $x_o=0.14$ for $J/t=0.4$. 
The experimentally obtained universal ratios for $Bi_2Sr_2CaCu_2O_{8+x}$, $La_{2-x}Sr_xCuO_4$ and $Tl_2Ba_2CuO_6$ are denoted as solid circles[20].
}
\end{figure}
 \end{minipage}

%%%%%%%%%%%%%%%%%%%%%%%%%%%%%%%%%%%%%%%%%%%%%%%%%%%%%%%%%%%%%%%%

\end{multicols}
\end{document}